\documentstyle[aps,preprint,tighten]{revtex}
 \input epsf.sty
 \input epsf.tex
 \input epsfig.sty
 
\def\pinax(#1,#2,#3,#4){\left(\matrix{#1 & #2\cr 
                                      #3 & #4\cr}\right)}
\def\tria(#1,#2,#3,#4,#5,#6,#7,#8,#9)
      {\left(\matrix{#1 & #2 & #3 \cr 
      #4 & #5 & #6 \cr 
      #7 & #8 & #9 \cr}\right)}
\def\bra(#1,#2){\left(\matrix{#1 \cr #2 \cr}\right)}
\def\ket(#1,#2){\left(\matrix{#1 & #2 \cr}\right)}
\def\P{{\cal P}}
\def\Re{\mbox{Re}}

\def\tr{\mbox{Tr}}
\def\nnn{\nonumber}

\def\bea{\begin{eqnarray}}
\def\eea{\end{eqnarray}}
\def\be{\begin{equation}}
\def\ee{\end{equation}}

\def\fact(#1,#2){\scriptstyle{#1\choose #2}}
\def\O{\Omega}
\def\pmb#1{\setbox0=\hbox{$#1$}%
	  \kern-.025em\copy0\kern-\wd0
	  \kern.05em\copy0\kern-\wd0
	  \kern-.025em\raise.0433em\box0 }

\begin{document}  

\title {\bf Abelian dominance and adjoint sources in lattice QCD}

\author{Grigorios I. Poulis}

\address{National Institute for Nuclear Physics and
 High-Energy Physics (NIKHEF)\\
P.O. Box 41882, NL-1009 DB Amsterdam, the Netherlands \\[2ex]}
\medskip
\date{January 1996}
\preprint{NIKHEF 95-064}

\maketitle

\begin{abstract}{\em }

Certain properties of maximal abelian projection are derived
which suggest that the fundamental and adjoint SU(2) string tensions
are reproduced by singly and doubly charged abelian Wilson loops, 
respectively. Thus, abelian dominance, which has been 
observed for color sources (quarks) in the fundamental 
representation, can be extended to higher 
representations. Numerical evidence in support of this 
conjecture also for adjoint quarks is presented.
The difference between maximal abelian and local projections is 
elucidated and the role of 
non-Wilson-like terms in the effective abelian action is discussed.

\end{abstract}


\section{Introduction}

Understanding the confinement mechanism remains one of the most 
important topics in non-perturbative QCD. A minimalistic approach
is to search for a subset of degrees of freedom that can lead to
confinement. In that respect, considerable progress has been made
in the case of compact quantum electrodynamics (CQED) where both 
analytical and numerical studies suggest that confinement can be 
understood to arise via monopole condensation in a dual superconductor
picture~\cite{Banks,Pol,dGT}. The abelian projection of 't Hooft~\cite{Hooft} 
is an attempt to apply the same ideas to QCD by mapping SU(N) gauge 
theory onto its U(1)$^{N-1}$ largest abelian (Cartan) subgroup. This 
mapping is effected by partial gauge fixing that leaves only a residual 
U(1)$^{N-1}$ symmetry. In the original paper by 't Hooft this is done 
by choosing a gauge so that an adjoint operator $X$ is diagonalized,
$ X\rightarrow V X V^{-1}$ = diag$[\exp(i\phi_1),\exp(i\phi_2),\dots
	\exp(i\phi_N)]$, with the constraint $\sum_i\phi_i=0$.
This gauge fixing is only partial since $V$ is determined modulo
left multiplication by factors
\begin{equation}\label{e2}
	d = \mbox{diag}\Bigl( e^{i\alpha_1},e^{i\alpha_2},\dots
	e^{i\alpha_N}\Bigr) \ ,\quad \sum_i\alpha_i=0 \ .
\end{equation}
$\{d\}$ lives in the Cartan subgroup U(1)$^{N-1}$.
After the abelian projection, the diagonal components $-i [A^\mu]_{ii}$
of the (adjoint) gauge field transform under this residual gauge symmetry 
as $N$ {\it abelian} potentials (labelled ``photons'' by 't Hooft),
$-i [A^\mu]_{ii}$ $\rightarrow $ $- i [A^\mu]_{ii}$ 
$-\partial^\mu \alpha_i$, where $\sum_i  [A^\mu]_{ii} = 0$. 
The off-diagonal components $-i [A^\mu]_{ij}$, $i\ne j$, 
transform as $N(N-1)$ charged vector fields
(labelled ``gluons'') $-i [A^\mu]_{ij} $ $\rightarrow $ 
$-i \exp(i[\alpha_i-\alpha_j])[A^\mu]_{ij}$.
Quark fields in the fundamental representation transform as singly charged
with respect to the appropriate ``photon'', $\psi_i$ $\rightarrow$ 
$ \exp(i\alpha_i) \psi_i$. The case of SU(2) is somehow special since the 
constraints imply that there is only one free phase $\alpha$ = 
$\alpha_1$ = $-\alpha_2$, with respect to which the off-diagonal ``gluons'' 
are {\it doubly} charged. Lattice QCD provides an ideal framework to 
carry out 't Hooft's programme~\cite{KronNP}.
On the lattice, an SU(2) 
link $U_{x,\mu}$ = $u_0 + i \vec\sigma\cdot\vec u$, with $u_0^2+\vec u^2=1$,
can be alternatively parametrized~\cite{Misha1,Misha2} as
$U^{11}_{x,\mu}$ = $\cos\phi_{x,\mu} 
\exp(i\theta_{x,\mu})$,  $U^{12}_{x,\mu}$ = $\sin\phi_{x,\mu} 
\exp(i\chi_{x,\mu})$,  $U^{21}_{x,\mu}$ = $-U^{12*}_{x,\mu}$,
$U^{22}_{x,\mu}$ =$U^{11*}_{x,\mu}$, with $\phi\in [0,\pi/2]$,
and $\chi,\theta\in (-\pi,\pi]$. The link $U_{x,\mu}$ can be decomposed 
as~\cite{KronPL,Misha1}
\begin{equation}\label{e6}
    U = \pinax(  \cos\phi\; e^{i\theta} , \sin\phi\; e^{i\chi},
      -\sin\phi \;e^{-i\chi}, \cos\phi\; e^{-i\theta} ) = 
      \pinax(  \cos\phi , \sin\phi\; e^{i\gamma},
      -\sin\phi \;e^{-i\gamma}, \cos\phi )
       \pinax( e^{i\theta},0,0,e^{-i\theta}) \ ,
\end{equation}
where the abelian phases $\theta$ are defined as
\begin{equation}\label{e7}
\theta \equiv \arctan\left( u_3\over u_0\right) \ ,
\end{equation}
and where $\cos\phi$ = $\sqrt{u_1^2+u_2^2}$. Under the residual U(1) 
symmetry $\theta$ and $\gamma=\chi+\theta$ transform like
\bea\label{b1}
\theta_{x,\mu} &\rightarrow & \theta_{x,\mu} + \alpha_x - 
\alpha_{x+\hat\mu} \nnn\\
\gamma_{x,\mu} &\rightarrow & \gamma_{x,\mu} + 2\alpha_x \ ,
\eea
that is, like abelian gauge field and charge-two matter field
(in the continuum), respectively. Notice that $\exp(i\theta)$ = 
$(u_0+iu_3)/\sqrt{u_0^2+u_3^2}$ and therefore 
diag$[\exp(i\theta),\exp(-i\theta)]$ can be viewed
as a rescaled diagonal SU(2) link. 
Besides photons, gluons and quarks the abelian projected theory
also contains abelian monopole world lines in four dimensions ($d=4$)
and monopole points (``instantons'') in $d=3$, which are identified
as singularities in the gauge-fixing condition and are on the lattice 
extracted from the 
phases $\theta$ following the algorithm of DeGrand and Toussaint~\cite{dGT}. 
Having started from QCD, one is then in position to repeat the numerical 
studies that have been done in the case of compact QED.
The abelian projection is gauge-dependent; the subset of degrees of 
freedom that can account for confinement may be different in different 
gauges~\cite{Misha1,Misha2} and choosing a gauge becomes an art, although 
there is some evidence that this gauge dependence reflects short distance 
fluctuations and tends to go away in the infrared~\cite{PTW}. 
Most lattice studies of the abelian projection are performed using 
the so called maximal abelian~\cite{KronPL} (MA) projection~\cite{term},
 corresponding in the continuum to
$D^\mu_0 A^\mu_{\pm}$ = $\partial^\mu  A^\mu_{\pm}$ $-$ 
$ig[A^\mu_0,A^\mu_{\pm}]$ = 0, where $A_{\pm}$ and $A_0$ are off diagonal
and diagonal gluons, respectively. On the lattice, with SU(2) gauge group, 
this amounts to making the SU(2) links maximally-diagonal after the 
abelian projection. This projection has  nice properties that seem to 
support 't Hooft's conjecture: the abelian monopole density in MA projection 
seems to scale~\cite{PTW,Born4d,Debbio4d,Born3d}; monopoles show a correlation with 
confinement in that they are dynamical in the confining phase and static 
above the finite temperature phase transition~\cite{KronNP,abeldom}. 
A particularly interesting feature of MA projection is that {\it abelian} 
Wilson loops constructed from the diagonal photons $\theta$ 
reproduce the fundamental QCD string tension~\cite{abeldom,Stack,Ken93}, 
a result named {\it abelian dominance}. This is an important result because
it is the photons $\theta$ that contain the monopoles that are supposedly 
responsible for confinement. Further support to the dual superconductor
picture is provided by measuring the string tension generated from monopoles 
alone and showing that it reproduces the full SU(2) string 
tension~\cite{Stack}. 

Abelian dominance may be interpreted as evidence that in maximal 
abelian (MA) projection the off-diagonal 
gluons $A_{\pm}$ can be neglected altogether, so that the Wilson loop is 
constructed simply from diagonal ``photons''. However, this ``diagonal
approximation''~\cite{Pol,Smit} is easily seen to be oversimplified by 
considering color sources 
(quarks) in SU(2) representations $j$ other than the fundamental ($j=1/2$). 
For $j=1$ (adjoint representation) the diagonal approximation 
forces the string tension to be zero not only in MA, but in  {\it any}
abelian projection, since  the $m_j=0$ 
component of the adjoint source cannot couple to diagonal gluons and remains 
unconfined~\cite{DD2,DGH}. This result is in disagreement with the 
fact that the adjoint potential shows for intermediate separations a
Casimir scaling linear behavior~\cite{Michael92,T95,PT,Mawh}
and has been interpreted as evidence for the failure of the
abelian projection mechanism~\cite{DD2,DGH,DD1}. 
Simulating QCD with adjoint color sources can therefore provide 
valuable insight into the dynamics of abelian projection. 

In this work the maximal
abelian projection is reviewed. 
It is shown that the dynamics of the theory in MA projection 
can be understood in terms of the following properties: 
(a) $<\cos\phi>$ is fixed to a value close to (but not exactly equal to) one 
(b) $\phi$ dependence factorizes
(c) the effective action  may be reasonably approximated 
    as involving only $\theta$ fields. At $\beta=2.4$, where the
calculations reported here are performed, $90\%$ of the SU(2) action is
carried by a CQED type action $\tilde\beta\cos\theta_P$, where
$\theta_P$ is an abelian plaquette constructed out of $\theta$ fields
and $\tilde\beta=\beta(\cos\phi)^4$ is an effective coupling.
There is also a ${\cal O}[10\%]$ 
correction that involves $\chi$-dependent terms and $\theta$-dependent 
terms coming from the expansion of the Fadeev-Popov determinant. 
Thus, to a first approximation the phases $\chi$ can be treated 
as random, and the phases $\phi$ as frozen.
By integrating over  $\chi$ and using $\phi$-factorization
it is shown that abelian dominance for sources in the fundamental
representation readily follows from this approximation. For adjoint sources 
the Wilson loop can be  split~\cite{DD1} into contributions from the
electrically (i.e., with respect to the residual U(1) symmetry)
 neutral ($m=0$) and electrically charged  ($m=\pm 1$) 
components of the adjoint source. Is is shown that the ``diagonal'' 
approximation is not reproducing the MA projection results as can
be most clearly seen by comparing their respective predictions for the 
electrically neutral  adjoint Wilson loop $W^{0}_{j=1}$: the diagonal 
approximation predicts it should be $1/3$, while ignoring the $\chi$-dependent
terms in the action predicts it should fluctuate around 0, albeit with a 
perimeter law falloff. This latter result simply means that to account for
the confining behavior of the $m=0$ components of the adjoint source, 
fluctuations of the diagonal gluon field have to be taken into account as well. 
On the other hand, the above approximation (i.e. ignoring the small, 
$\chi$-dependent terms in the action) implies that the string tension 
from the electrically charged  part of the adjoint Wilson loop $W^{\pm}_{j=1}$ 
must be given by doubly charged abelian Wilson loops $W_{n=2}$. 
It is numerically found that  $W^{\pm}_{j=1}$ generates roughly the
same string tension as the full adjoint Wilson loop $W_{j=1}$ = 
$W^{\pm}_{j=1}+W^{0}_{j=1}$, in accord with calculations in 
three dimensions~\cite{DD1}. This last result may then be used as
phenomenological input, which, combined with the above analysis,
suggests that doubly charged abelian Wilson loops
should reproduce the full SU(2) string tension in MA projection.
Numerical evidence in support of this result is presented.
Thus, abelian dominance can be extended to quarks transforming
in the adjoint representation. Doubly charged loops in CQED generate a string 
tension that Casimir-scales, and is therefore equal to $2^2=4$ times 
the one from singly charged loops~\cite{TW}. 
Abelian dominance for adjoint sources on the other hand means that this ratio is
in MA projection equal to the ratio of quadratic Casimirs $j(j+1)$ between
 adjoint ($j=1$) and fundamental $(j=1/2$) SU(2) representations, that is,
$8/3$ $\approx$ 2.7 (in F12 projection is found to be
less than 2). These results show that abelian QCD is certainly closer to 
CQED in MA projection than it is in local projections like F12
but still quite different, as is necessary due to the qualitative differences
such as deconfinement phase transition, asymptotic freedom etc.,
between QCD (which abelian QCD in MA projection presumably reproduces) and CQED.
In the context of the MA projection properties we mentioned above these
non-Wilson type of $\theta$-dependent action terms originate in the
expansion of the Fadeev-Popov determinant. Finally, the mechanism of charge 
screening (breaking of the adjoint flux tube at large quark-antiquark
separations) is discussed in the framework of MA projection abelian QCD. 

The structure of this article is as follows: in section II 
we discuss some properties of MA projection and then use them to derive
abelian dominance for fundamental quarks (in section III) and  for
adjoint quarks (in section IV). A summary of our investigation is 
presented in section V.

\section{The Maximal Abelian projection}

\begin{figure}[htb]
\begin{center}
\mbox{\epsfig{file=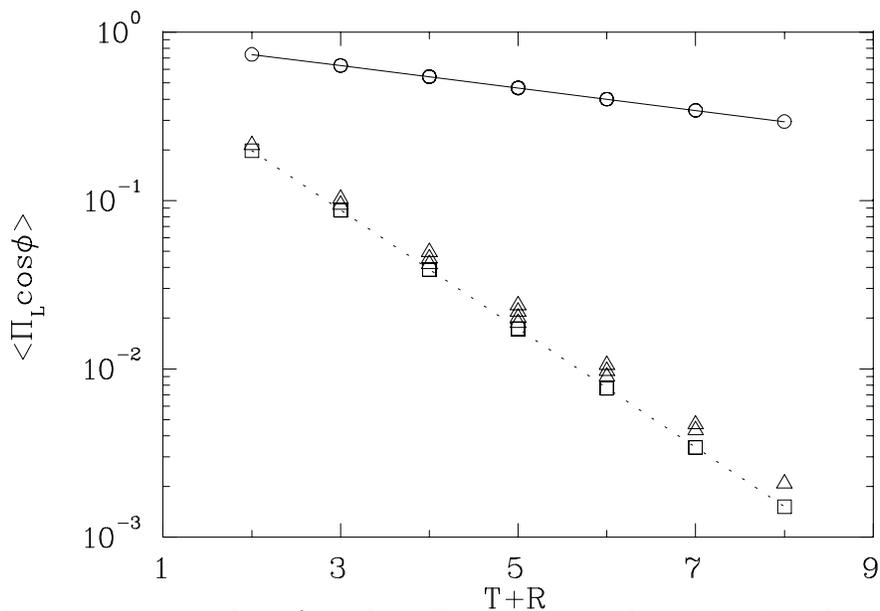,width=.70\textwidth,angle=90}}
\caption{ The expectation value of product $\prod_L\cos\phi$ 
around an $T\times R$ Wilson loop. Results are shown for links 
unprojected ($\Box$), F12-projected ($\triangle$)
and MA-projected (o) QCD. The solid curve corresponds
to  $0.9263^{\P}$ and the dotted curve to $(2/3)^{\P}$, where
$2/3$ is the expectation value of randomly distributed $\cos\phi$
and ${\P}$ the loop perimeter, ${\P}=2(R+T)$. }
\label{cos}
\end{center}
\end{figure}

\begin{figure}[htb]
\begin{center}
\mbox{\epsfig{file=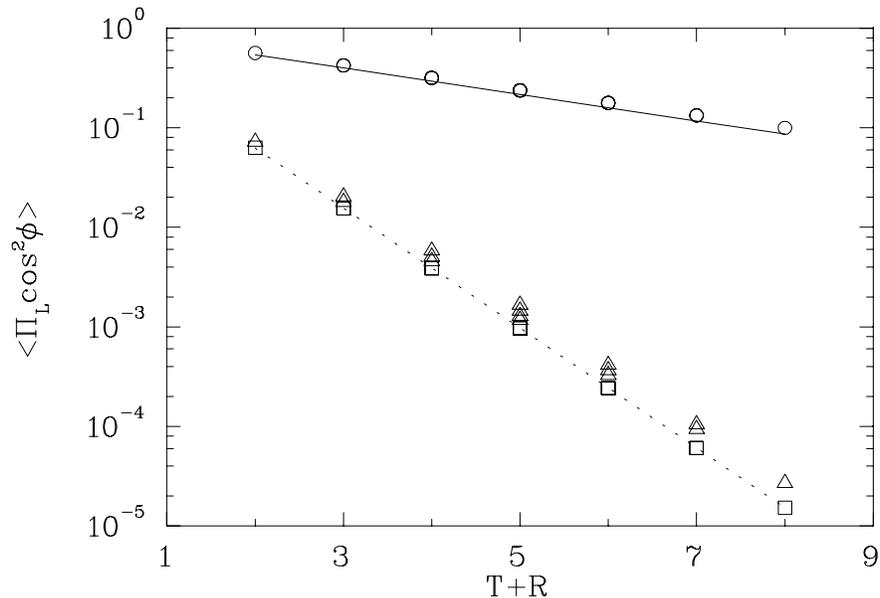,width=.70\textwidth,angle=90}}
\caption[dummy]{ Same as in Fig.\ \ref{cos}, for the expectation value of 
$\prod_L\cos^2\phi$. The solid curve corresponds
to  $0.9263^{2\P}$ and the dotted curve to $(1/2)^{\P}$, where
$1/2$ is the expectation value of randomly distributed $\cos^2\phi$.}
\label{cos2}
\end{center}
\end{figure}

MA projection amounts to maximizing the quantity 
\begin{equation}\label{om}
\Omega\equiv \sum_{x,\mu}\rm{Tr}\left[ U_\mu^\dagger(x)\sigma_3
U_\mu(x)\sigma_3\right] =  \sum_{x,\mu}\cos(2\phi_{x,\mu}) \ .
\end{equation}
Following Ref.~\cite{Misha1} the SU(2) plaquette action $S_P$
is decomposed in $S_P=S_\theta + S_\chi + S_{\theta\chi}$.
$S_\theta$ contains $\theta$ but not $\chi$ fields, $S_{\chi}$
involves $\chi$ but not $\theta$ fields, and $S_{\theta\chi}$ 
involves interaction between  $\theta$ and $\chi$ fields:
\be\label{sp}
S_P = {1\over 2}\tr \left( U_1\; U_2\; U_3^\dagger\; U_4^\dagger\right) = 
      S_\theta + S_\chi + S_{\theta\chi} \ ,
\ee
where
\bea\label{sigmas}
 S_\theta &=& \Bigl(\cos\phi_1\cos\phi_2\cos\phi_3\cos\phi_4\Bigr)
\cos\theta_P\\
 S_\chi &=& \Bigl(\sin\phi_1\sin\phi_2\sin\phi_3\sin\phi_4\Bigr)\cos\chi_P\\
 S_{\theta\chi} &=&
-\Bigl(\cos\phi_1\cos\phi_2\sin\phi_3\sin\phi_4\Bigr)\cos(\theta_1+\theta_2
+\chi_3-\chi_4) + \dots 
\eea
Here $\theta_P \equiv \theta_1+\theta_2-\theta_3-\theta_4$  
and $\chi_P \equiv \chi_1-\chi_2+\chi_3-\chi_4$  are U(1)-gauge invariant
plaquettes constructed from phases $\theta$ and $\chi$, respectively.
The dots stand for 5 more terms involving U(1) invariant (c.f. 
Eq.~(\ref{b1})) combinations of two $\theta$ and two $\chi$ phases each. 
$S_\theta$ is proportional to the Wilson action for 
compact electrodynamics. The gauge fixing condition, Eq.~(\ref{om}),
forces the $\phi$ phases to fluctuate around zero and thus one 
expects~\cite{Misha1} that $S_\chi <  S_{\theta\chi} <S_\theta $.
Using 50 configurations on a $12^4$ lattice at $\beta=2.4$ we find 
that $<\cos\phi_{x,\mu}>$ is equal to
0.6665(1), 0.6711(1) and 0.9263(1) in unprojected, F12-projected, and 
MA-projected QCD, respectively. The value in the
former two cases corresponds to $\phi$ being basically random, as can be seen 
using the group measure corresponding to link parametrization, Eq.~(\ref{e6}) 
\be\label{measure}
{1\over 2 \pi^2} \int_{0}^{\pi/2}d\phi \cos\phi\sin\phi
\int_{-\pi}^{\pi}d\theta\int_{-\pi}^{\pi} d\chi \ ,
\ee
from which follows that $<\cos\phi>$ = $2/3$.
In MA projection $<\cos\phi>$ is close to 1 and, accordingly, we find
$S_P$ = 0.6298(1), $S_\theta$ = .5739(4) which means that they
differ by $(S_P-S_\theta) / S_P$ = 9\%. Moreover we find that 
$<\prod_{P}\cos\phi><\cos\theta_P>$ =0.5586(5)  which shows that 
the SU(2) plaquette to a good approximation factorizes in a 
product of $\cos\phi$ terms
around the loop times  an abelian plaquette constructed from
the photons $\theta$. In F12 projection we find that $S_\theta$ = 0.0890(1)
accounts for a small fraction of $S_P$ and, moreover, that factorization 
does not hold as  $<\prod_{P}\cos\phi><\cos\theta_P>$ = 0.0565(1).
In MA projection $\cos\phi$  behaves like a parameter
(more appropriately, like a spin-glass coupling since the
effective coupling is $\beta\cos\phi^4$); for example,
$<\prod_{P}\cos\phi>$ = 0.7377(3), while $<\cos\phi>^4$ = 0.7362.
This behavior is shown in Fig.~\ref{cos} for the product 
$<\prod_{L}\cos\phi>$ around $T\times R$ rectangular Wilson loops. 
The solid curve
corresponds to  $0.9263^{\P}$
and the dotted curve to $(2/3)^{\P}$, where $\P=2(R+T)$
is the perimeter of the loop. It is clear
that the product $<\prod_{L}\cos\phi>$ around Wilson loops falls with the
perimeter of the loop. In F12 projection $<\cos\phi>$ is slightly larger
than in the unprojected case because of the gauge condition (diagonalization
of the plaquette in the (1,2) plane). For future reference we show
the behavior of $<\prod_{L}\cos^2\phi>$ in Fig.~\ref{cos2}.
The solid curve corresponds to $0.9263^{2\P}$
and the dotted curve to $(1/2)^{\P}$, where $1/2$ 
is, from Eq.~(\ref{measure}), the expectation value for 
randomly distributed $\cos^2\phi$.

\section{Abelian Dominance: fundamental sources}

\begin{figure}[htb]
\begin{center}
\mbox{\epsfig{file=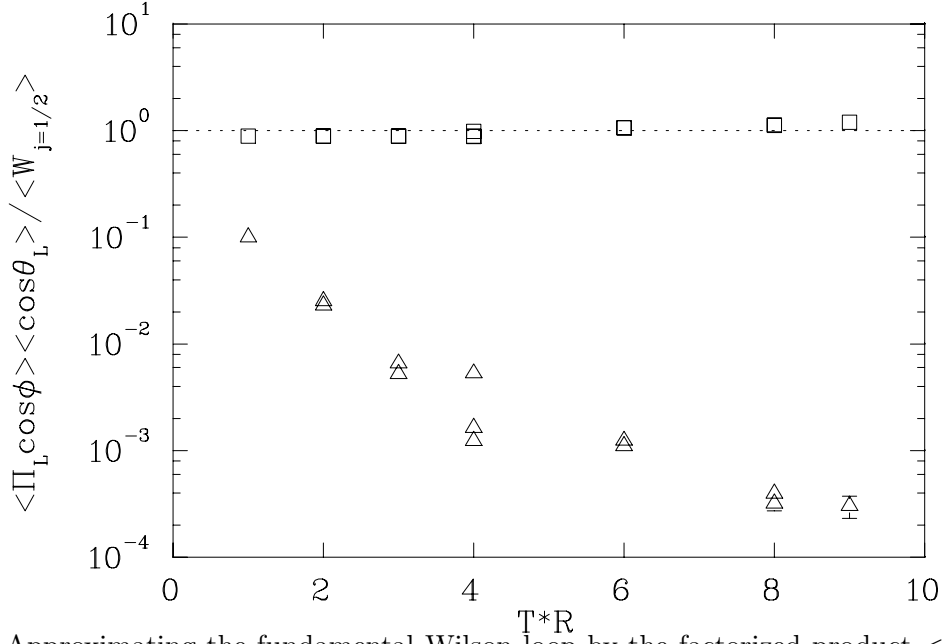,width=.75\textwidth,angle=90}}
\caption{ Approximating the fundamental Wilson loop by the
factorized product $<\prod_L\cos\phi>$ around the $T\times R$ loop
times the singly charged abelian Wilson loop $<\cos\theta_L>$.
Results shown in F12 ($\triangle$) and MA ($\Box$) projection.}
\label{perc}
\end{center}
\end{figure}

\begin{figure}[htb]
\begin{center}
\mbox{\epsfig{file=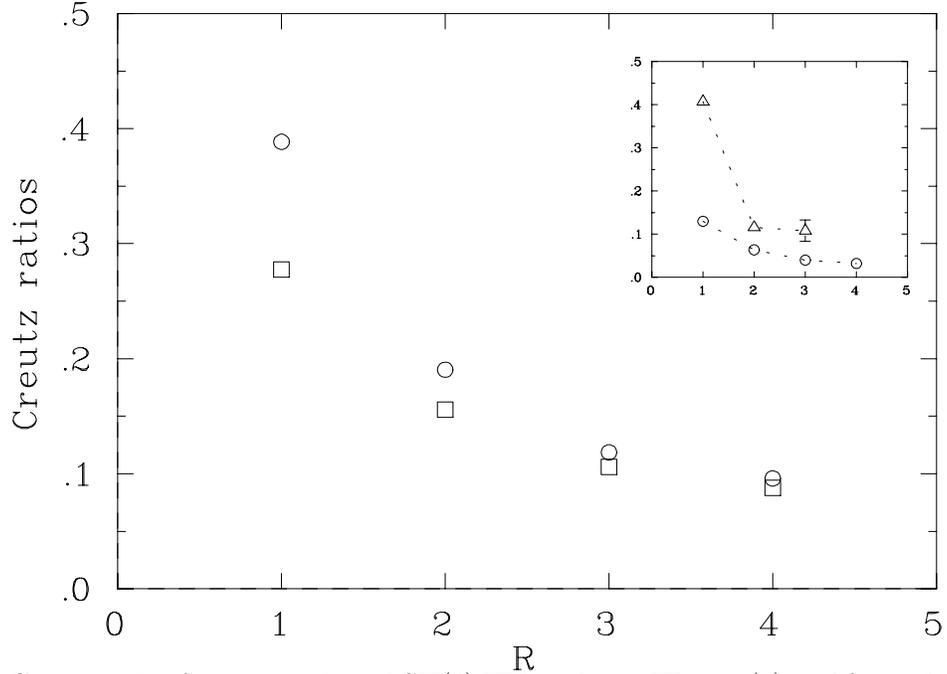,width=.75\textwidth,angle=90}}
\caption{	Creutz ratios from unprojected SU(2) Wilson loops 
        $W_{j=1/2}$  (o) and from abelian Wilson loops
        $W_{n=1}$ (equivalently, $W^d_{j=1/2}$) 
	in  F12 ($\triangle$) and MA ($\Box$) projection.}
\label{fad}
\end{center}
\end{figure}

For an $T\times R$ Wilson loop in the fundamental representation, 
$W$ = $w_0+i\vec \sigma\cdot\vec w$, we can write a generalization of
Eq.~(\ref{sigmas}),
\bea\label{w03}
w_0 + i w_3 &=& (\cos\phi)^{2L}\exp(i\theta_L) + 
	      (\sin\phi)^{2L}\exp(i\chi_L) \nnn\\
           && +   \sum_{m=1}^{L-1}(\cos\phi)^{2m}(\sin\phi)^{2(L-m)}
              \sum_{n=1}^{\fact(2L,2m)} s_n 
             \exp\Bigl(i\O_n[2m\theta,2(L-m)\chi]\Bigr) \ .
\eea
The notation is as follows: $L$ stands for $R+T$, and $\theta_L$
is the abelian, singly charged, Wilson loop constructed from photons
$\theta$. Similarly, $\chi_L$ is the charged matter field analog of Wilson
loop and  $\O_n[2m\theta,2(L-m)\chi]$ is an angle containing $2m$ 
$\theta$-phases and $2(L-m)$ $\chi$-phases. 
There can be an even-only number of 
either of them and therefore $m=1$ for the $1\times 1$ Wilson loop.
The index $n$ denotes the specific combination of phases once $m$ is
fixed. There are $\fact(2L,2m)$ such combinations for given $L$ and $m$, 
so that the total number of terms is $2+\sum_{m=1}^{L-1}
\fact(2L,2m)$ = $2^{2L-1}$ (for example, in the case of the plaquette
R=T=1, L=2, m=1, and the total number of terms is 8). 
Finally, $s_n$ is a sign factor for the given phase combination.
Henceforth, $\cos\phi$ and $\sin\phi$ are treated as fixed parameters,
that is, integration over $\phi$ using the MA projection condition, 
Eq.~(\ref{om}),
amounts to fixing the magnitude of $\cos\phi$ with the residual effect of
corrections to the action introduced by the Fadeev-Popov 
determinant~\cite{Misha1}.  
Consider now the expectation value of the trace of the fundamental Wilson
loop, $<W_{j=1/2}>$ = $<w_0>$. Ignoring the small $\chi$-dependent part in the
action simply means that all terms in the operator involving angles 
containing $\chi$ variables vanish and therefore
\bea\label{pol1}
<W_{j=1/2}> &\approx& (\cos\phi)^{2L}\int d[\theta] e^{-[\beta S_\theta + \Delta
S_{FP}] }\cos\theta_L \nnn\\ 
&=& (\cos\phi)^{2L} <\cos\theta_L>_{abel} \ .
\eea
Here $\Delta S_{FP}$ is an effective action coming from the Fadeev-Popov
determinant after $\phi$ integration.
From Fig.~\ref{perc} one sees that  Eq.~(\ref{pol1}) is a good 
approximation in MA projection 
and, not surprisingly, very bad in F12 projection. Abelian dominance 
follows  from Eq.~(\ref{pol1}):  the string tensions 
from the SU(2) expectation value $<W_{j=1/2}>$ and the abelian
expectation value $<\cos\theta_L>_{abel}$ (that is, the singly charged abelian
Wilson loop) should be equal since the two differ by a 
perimeter term that disappears when forming Creutz ratios. Notice, for
future reference, that $\cos\theta_L$ can be thought as constructed from
{\it either} U(1) links $\exp(i\theta)$, Eq.\ (\ref{e7}), 
oriented along a path $L$
\begin{equation}\label{e10}
	W_{n=1}= \Re \left\{ \prod_{i\in L} e^{i\theta_i} \right\}
                = \cos\Bigl( \sum_{i\in L} \theta_i\Bigr) 
               \equiv \cos\bigl(\theta_L\bigr) \ ,
\end{equation}
{\it or}, from rescaled diagonal SU(2) links
\begin{equation}\label{e11}
 W^{d}_{j=1/2}={1\over 2}
             \tr\left\{  \prod_{i\in L}
                  \pinax( e^{i\theta_i},0,0,e^{-i\theta_i})\right\}
             = \cos\Bigl( \sum_{i\in L} \theta_i\Bigr) = W_{n=1} \ .
\end{equation}
Abelian dominance is by now well established~\cite{abeldom,Stack,Ken93}. 
For completeness however, we test abelian dominance for the fundamental
 case in Fig. \ref{fad}. The
Creutz ratios from full SU(2) Wilson loops $W_{j=1/2}$
are compared to Creutz ratios from singly charged abelian Wilson loops
$W^{AP}_{n=1}$ (equivalently, fundamental diagonal (rescaled) SU(2) Wilson 
loops $W^{d,AP}_{j=1/2}$) in maximal abelian (AP=MA) and field strength 
(AP=F12) projection. Calculations are performed at $\beta=2.4$, 
in four dimensions, where 
monopoles in MA projection seem to scale~\cite{Born4d,Debbio4d} and
fundamental abelian dominance has been already observed~\cite{Suz92}.
Results for SU(2) Creutz ratios in fundamental and adjoint representation
come from 80 measurements on a $16^4$ lattice separated by 20 updates. 
An iterative smearing (with 20 iterations) is used for the space-like links, 
with parameter $c=2.5$~\cite{fuzz}. The abelian projection results come 
from two sets of measurements. The first comprises of 200 measurements on a
$16^4$ lattice and the second of 350 measurements on a $12^4$ lattice. 
Measurements are separated by 60 updates. Error bars for the effective
potential and Creutz ratios are obtained using the jackknife method.
The MA condition, Eq.~(\ref{om}), is enforced iteratively using the 
overrelaxation algorithm of 
Ref.\ \cite{Mand} with parameter $\omega=1.7$. 
Since the effective U(1) theory after projection fixing is not precisely 
known one cannot use multihit~\cite{multihit} 
variance reduction techniques. A straightforward modification of smearing 
was tried by creating smeared U(1) links after the abelian projection 
which did not prove successful. Without such techniques the 
range of $R,T$ values that results can be obtained becomes severely restricted, 
especially for the operators relevant for the adjoint source case (next section).
No attempt is made to fit the Creutz ratios to some ansatz from which to 
extract the string tension.
In MA projection abelian dominance {\it is} observed as the full SU(2) Creutz ratios are 
reproduced by the abelian ones at intermediate separations $R$
\begin{equation}\label{ne1}
	\sigma_{j=1/2} = \sigma^{MA}_{n=1} \ .
\end{equation}
Here, $n$ labels the U(1) representation (charge) in the abelian Wilson loop
operator, $W_n = \cos(n\theta_L)$. Fundamental quarks transform
as singly charged with respect to the residual U(1), hence $n=1$ in this case. 
The fact that Creutz ratios show more noise in F12 projection than in MA projection
reflects short distance fluctuations that plague local projections such as F12
but which are washed out in maximal abelian projection which is non-local~\cite{PTW}.
F12 projection does not satisfy abelian dominance: the abelian string tension is
higher than the SU(2) one~\cite{Ken93} in commensurate with higher monopole density
in local projections like F12 compared to MA projection~\cite{Ken93,PTW}.

\section{Abelian Dominance: adjoint sources}

\begin{figure}[htb]
\begin{center}
\mbox{\epsfig{file=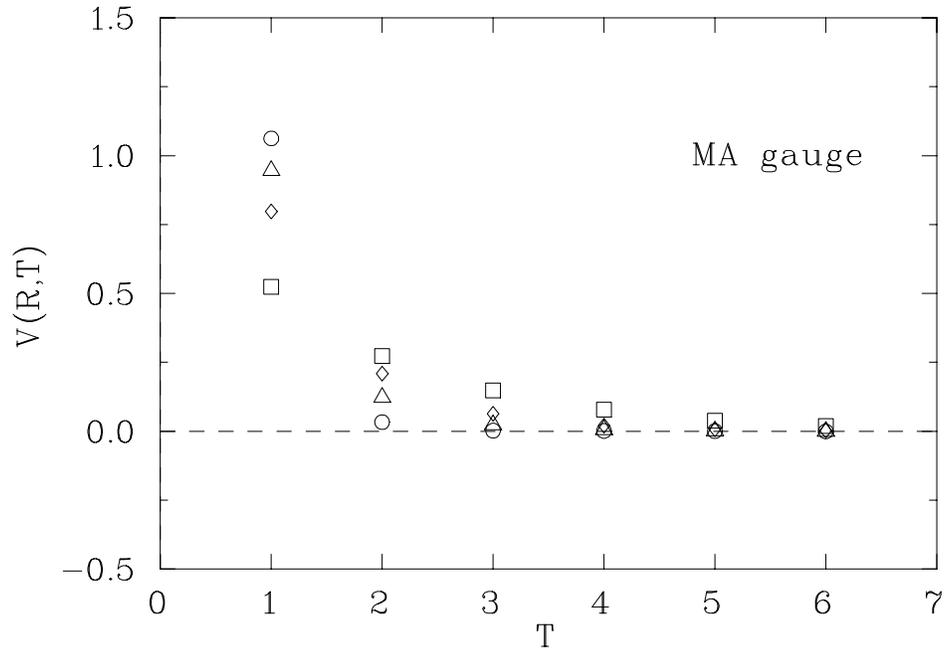,width=.75\textwidth,angle=90}}
\caption{       Time-dependent  potential $V(R,T)$ 
        from adjoint diagonal SU(2) Wilson loops $W^{d}_{j=1}$ 
        in MA projection, for $R = 1$ ($\Box$), 2 
        ($\diamond$), 3 ($\triangle$), and 5 (o). }
\label{ama}
\end{center}
\end{figure}

\begin{figure}[htb]
\begin{center}
\mbox{\epsfig{file=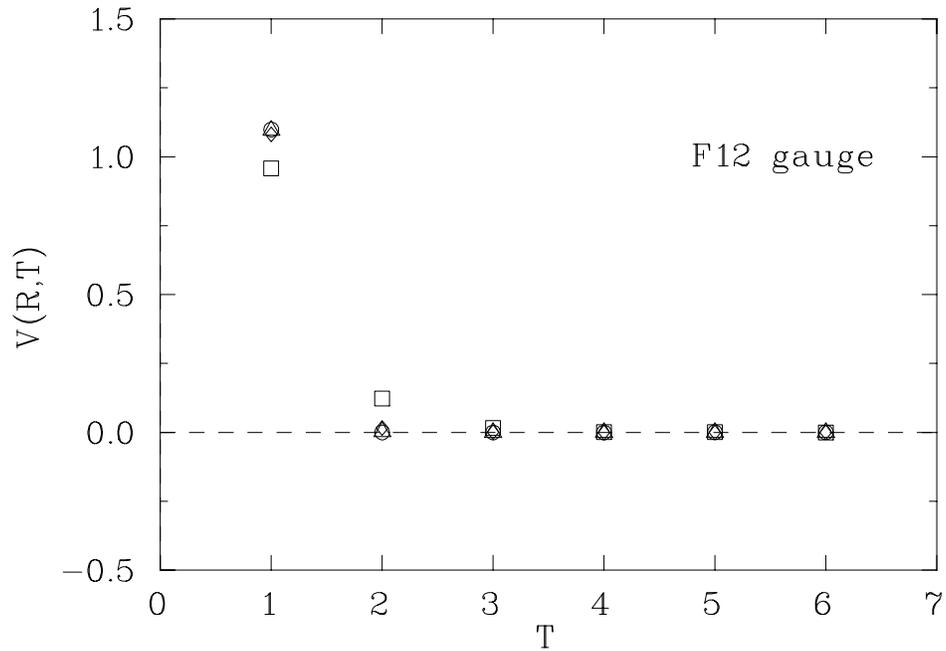,width=.75\textwidth,angle=90}}
\caption[dummy]{ Same as in Fig.\ \ref{ama} but in F12 projection.}
\label{af}
\end{center}
\end{figure}

\begin{figure}[htb]
\begin{center}
\mbox{\epsfig{file=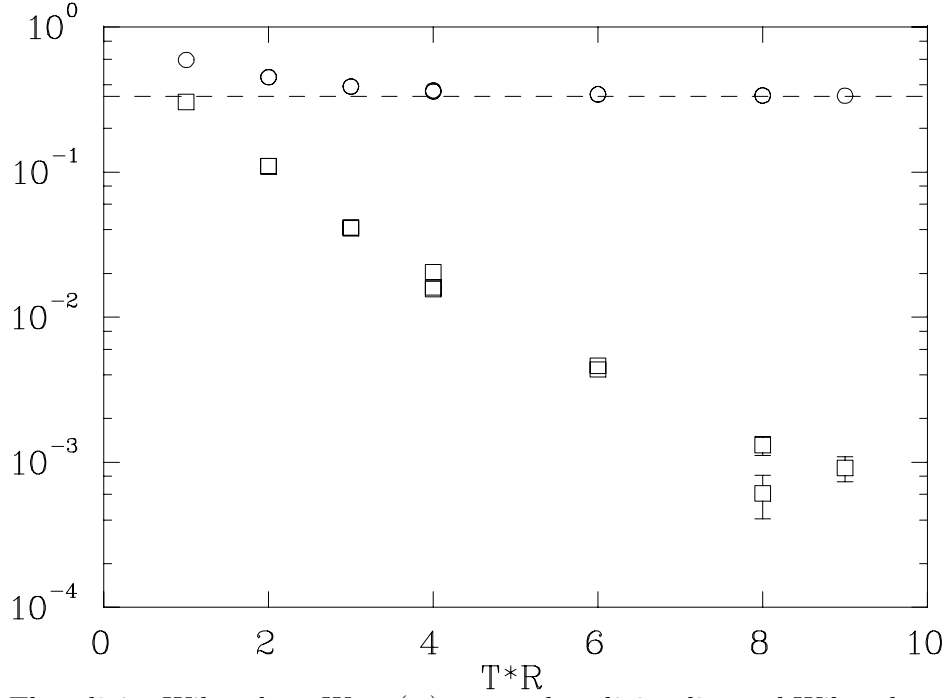,width=.75\textwidth,angle=90}}
\caption{ The adjoint Wilson loop $W_{j=1}$ ($\Box$) versus the
adjoint diagonal Wilson loop $W^d_{j=1}$ (o) in MA projection.
The dashed line corresponds to the asymptotic value for the latter, 
$W^d_{j=1}=1/3$.}
\label{www}
\end{center}
\end{figure}

Consider now color sources (quarks) transforming in the adjoint 
representation of SU(2). At intermediate separations $R$ the potential 
between adjoint sources shows a linear behavior with a string tension 
that exhibits Casimir scaling, that is, scales like $j(j+1)$, where $j$ labels
the SU(2) representation ($j=1/2$ for the fundamental, $j=1$ for the adjoint, 
etc.) Thus the adjoint string tension is roughly $8/3$ times the fundamental
one. Eventually however, the adjoint potential is 
expected to saturate as it becomes energetically favorable for color-singlet 
bound states of adjoint source and glue (``gluelumps'') to screen the 
adjoint flux tube~\cite{Pol,Zou,Michael85}. 
The Casimir scaling behavior is of perturbative (weak coupling) 
origin~\cite{BK}: the one-gluon exchange contribution to the potential
scales with the charge squared, hence the  Casimir factor $j(j+1)$.
There is substantial numerical evidence in favor of Casimir scaling
in both three~\cite{PT,Mawh} and four~\cite{Michael92,T95} 
dimensions, for SU(N), as well as compact QED~\cite{TW} (where $n$-ply 
charged Wilson loops develop a string tension that scales like $n^2$). 
Surprisingly enough, Casimir scaling seems to persist at large 
separations. In three dimensions it is still seen at $R$ values 
corresponding to almost twice the screening distance where 
gluelumps were expected to have saturated the adjoint potential~\cite{PT}.
Thus, at least for separations $R\le 7a$ in $d=4$ at $\beta=2.4$ ($a\approx 0.12$
fm), where our measurements are made,
{\it adjoint abelian dominance} requires 
abelian QCD in maximal abelian projection to generate a string tension 
\begin{eqnarray}
\sigma^{abel, MA}_{j=1} &\stackrel{\mbox{?}}{=}& \sigma_{j=1} \label{aad1} \\
                        &=& {8/3}\;\sigma_{j=1/2}\label{aad2} \\
                        &=& {8/3}\;\sigma^{MA}_{n=1}\label{aad3} \ .
\end{eqnarray}
 where $8/3$ $\approx$ $2.7$ is the ratio of  Casimirs between adjoint  and 
fundamental SU(2) representations. Eq.\ (\ref{aad1}) would be {\it direct}
evidence for adjoint abelian dominance, whereas Eq.\ (\ref{aad3}) 
would be {\it indirect}, since it involves comparison between {\it abelian} 
observables in MA projection and is based on the observed Casimir scaling of
the adjoint string tension, Eq.\ (\ref{aad2}), and the established 
abelian dominance for fundamental sources, Eq.\ (\ref{ne1}).
The critical issue is to decide from which abelian Wilson loops should  
$\sigma^{abel, MA}_{j=1}$ be extracted. We shall come to this question
shortly.
The adjoint string tension $\sigma_{j=1}$ is obtained from
adjoint SU(2) Wilson loops, defined in a generic representation $j$ as~\cite{TW}
\begin{equation}\label{e14}
	W_j \equiv {1\over 2j+1} \mbox{Tr} \left\{
             \prod_{i\in{L}} {\cal D}_j[U_i] \right\} ,
\end{equation}
with ${\cal D}[U_i]$ the appropriate irreducible representation of the 
link $U_i$ $\in$ SU(2). The adjoint ($j=1$) representation of 
an SU(2) element  $U$ with fundamental representation 
${\cal D}_{1/2}[U]$ = $u_0 + i \vec\sigma\cdot\vec u$, with $u_0^2+\vec u^2=1$,
reads~\cite{infi}
\begin{eqnarray}\label{e12}
	\left( {\cal D}_{1}[U]\right)_{ab}
           &=& {1\over 2} \mbox{Tr}\Bigl( 
        \sigma_a {\cal D}_{1/2}[U]\sigma_b {\cal D}^\dagger_{1/2}[U] \Bigl) \\
        &=& \delta_{ab}(2 u_0^2 -1) + 2 u_a u_b + 2 \epsilon_{abc} u_0 u_c 
         \nnn \ ,      
\end{eqnarray}
where $a,b\in\{1,\dots N^2-1=3\}$. Using  Eqs.\ (\ref{e14},\ref{e12}) one has
\begin{equation}\label{e16a}
	W_{j=1} = {4 (W_{j=1/2})^2 - 1 \over 3} = {4 w_0^2 - 1 \over 3} \ .
\end{equation}
Writing $W$ = $w_0+i\vec \sigma\cdot\vec w$ for the
fundamental Wilson loop, the trace of the adjoint Wilson loop can be split
into the contribution of neutral ($m=0$) and charged ($m=\pm 1$) states~\cite{DD1}. 
From Eq.\ (\ref{e12}) we obtain
\begin{eqnarray}
W^{0}_{j=1} &\equiv& { \left({\cal D}_{1}[W]\right)_{33}\over 3} 
= {2(w_0^2+w_3^2)-1 \over 3} \label{wn} \\
W^{\pm}_{j=1} &\equiv &{ \left({\cal D}_{1}[W]\right)_{11}+\left(
{\cal D}_{1}[W]\right)_{22}\over 3} = {2\over 3}(w_0^2-w_3^2)\label{wch} \ .
\end{eqnarray}
This decomposition is projection-dependent as only the sum 
(\ref{wn}) $+$ (\ref{wch}) = (\ref{e16a}) is SU(2)-invariant. 
 Before applying the considerations
of the previous section to the operators (\ref{wn},\ref{wch}) let us
discuss the abelian projection in the approximation where 
the Wilson loop of the full SU(2) 
theory is dominated by diagonal gauge field contributions~\cite{Pol,Smit}
\begin{eqnarray}\label{e18}
           <W_j ({\cal C})> &\equiv& {1\over 2j+1} 
             < \mbox{Tr}\exp\left(i  \oint dx^\mu A^a_\mu T_a^j\right) >\nnn\\
&\approx& {1\over 2j+1} 
             < \mbox{Tr}\exp\left(i  \oint dx^\mu A^3_\mu T_3^j\right) >\nnn\\
&=& {1\over 2j+1} \sum_{m=-j}^j
             < <m|\exp\left(i  \oint dx^\mu A^3_\mu T_3^j\right)|m> > \nnn\\
&=& {1\over 2j+1} 
             \sum_{m=j}^j<\exp\left(i m \oint dx^\mu A^3_\mu \right)> \ .
\end{eqnarray}
Although Eq.\ (\ref{e18}) is sometimes referred to as the ``abelian dominance'' 
approximation~\cite{DD2}, we prefer the term ``diagonal approximation''. 
Notice that if there was a ``perfectly abelian'' projection, that is, one where
the coset fields would be identically zero, $u_2^2+u_1^2\equiv 0$, then
the above approximation would be {\it exact}. Abelian dominance would
{\it also} be exact, since the full SU(2) Wilson loop operator
would be identical to the one obtained from  diagonal rescaled SU(2) links
(equivalently, the singly charged abelian Wilson loop, Eq.~(\ref{e11})).
Of course such a projection does not exist, but due to its very definition 
maximal abelian projection seems to be a close approximation to the diagonal 
SU(2) limit, since (in the language of section II) the diagonal
approximation corresponds to $\cos\phi=1$. By just observing
abelian dominance for fundamental quarks one cannot distinguish 
between the diagonal approximation and the approximations of section II
since their difference affects only the perimeter term (c.f. Eq.~(\ref{pol1})). 
The reason for distinguishing between them becomes clear when introducing 
color sources (quarks) in higher than the fundamental representation.
In particular, consider a color source in the adjoint representation of SU(2) 
\begin{equation}\label{ee1}
\Phi \rightarrow G\;\Phi\; G^{-1} \ .
\end{equation}
Under the residual abelian symmetry $G$ = diag$(\exp(ia),
\exp(-ia))$ we write, using Eq.\ (\ref{e12})
\begin{equation}\label{kai}
({\cal D}_1[\Phi]) \rightarrow \tria(\cos(2 a), \sin(2a), 0 , -\sin(2a),
 \cos(2a) ,0  ,0, 0, 1)  ({\cal D}_1[\Phi]) \ .
\end{equation}
Thus, the $m=\pm 1$ source components (corresponding to indices 1,2 in 
the above matrix) are electrically doubly charged (that is, they interact
with the photons of the abelian projection) while the $m=0$ component
(corresponding to index 3) is electrically neutral and does not
interact with photons. In the diagonal approximation therefore, one expects
from  Eq.~(\ref{wn}) that $<W^0_{j=1}>$ $\rightarrow$ $1/3$, corresponding to
the $m=0$ contribution in Eq.~(\ref{e18}). Even if $<W^{\pm}_{j=1}>$ 
does develop a string tension, the total adjoint Wilson loop should
fluctuate around $1/3$ and, clearly, the adjoint string tension should be
zero. Thus, approximation (\ref{e18}) imposes dramatic constraints to higher 
representation string tensions: for {\it even} representations $j$, such as
the  adjoint, the $m=0$ eigenvalue contributes a factor of $(2j+1)^{-1}$ in 
$<W_j ({\cal C}) >$ which does not allow for confining behavior. 
One therefore expects~\cite{Pol,Smit} 
\begin{eqnarray}\label{e19a}
	j =\mbox{even} \quad &\Rightarrow& \quad \sigma_j = 0 \\
	j =\mbox{odd} \quad  &\Rightarrow& \quad \sigma_j = 
             \sigma_{1/2} \nnn\ .
\end{eqnarray}
The adjoint string tension is of course expected to be zero at separations
beyond the screening distance $R_c$. Still, though, the abelian projection mechanism
should be able to generate the observed Casimir scaling adjoint string tension 
for $R<R_c$. Within the diagonal approximation, the abelian (i.e,
constructed exclusively from $\theta$ links) operator
that should generate the adjoint SU(2) string tension is the adjoint
Wilson loop constructed from rescaled diagonal SU(2) links, diag$(\exp(i\theta),
\exp(-i\theta))$~\cite{DD2} 
\begin{equation}\label{e16}
	W^{d}_{j=1} = {4 (W^d_{j=1/2})^2 - 1 \over 3}
                    = {4\cos^2(\theta_L) - 1 \over 3} 
                    = {2\over 3}\cos(2\theta_L) + {1\over 3} \ .
\end{equation}
However, the discussion above suggests that this operator can {\it not} 
lead to a confining potential. Indeed, in Ref.\ \cite{DD2} it was
shown that the SU(2)  string tension  in three dimensions from adjoint 
diagonal Wilson loops in MA projection is zero. This is not so much
a test of the diagonal approximation, but rather, is a demonstration 
that Eq.\ (\ref{e19a}) {\it does}  follow from the diagonal approximation, 
{\it whether the latter is justified or not}. In order to make this point 
clear, we show the effective (time dependent) potential 
\begin{equation}\label{effpot}
V(R,T)^{d,AP}_{j=1}=
\log\left[ W^{d,AP}_{j=1}(R,T-1)\over W^{d,AP}_{j=1}(R,T) \right] \ .
\end{equation}
in four dimensions for maximal abelian projection (AP=MA) in Fig.\ \ref{ama}
and field strength projection (AP=F12) in Fig.\ \ref{af}.
The potential is consistent with zero in both projections; not only the
string tension vanishes but the perimeter term as well. This verifies
the expectation from Eqs.\ (\ref{e18},\ref{wn}) that in a diagonal SU(2) theory the 
adjoint Wilson loop $W^{d,AP}_{j=1}$ $\rightarrow$ $1/3$, as is evident from
Fig.~\ref{www}.

\begin{figure}[htb]
\begin{center}
\mbox{\epsfig{file=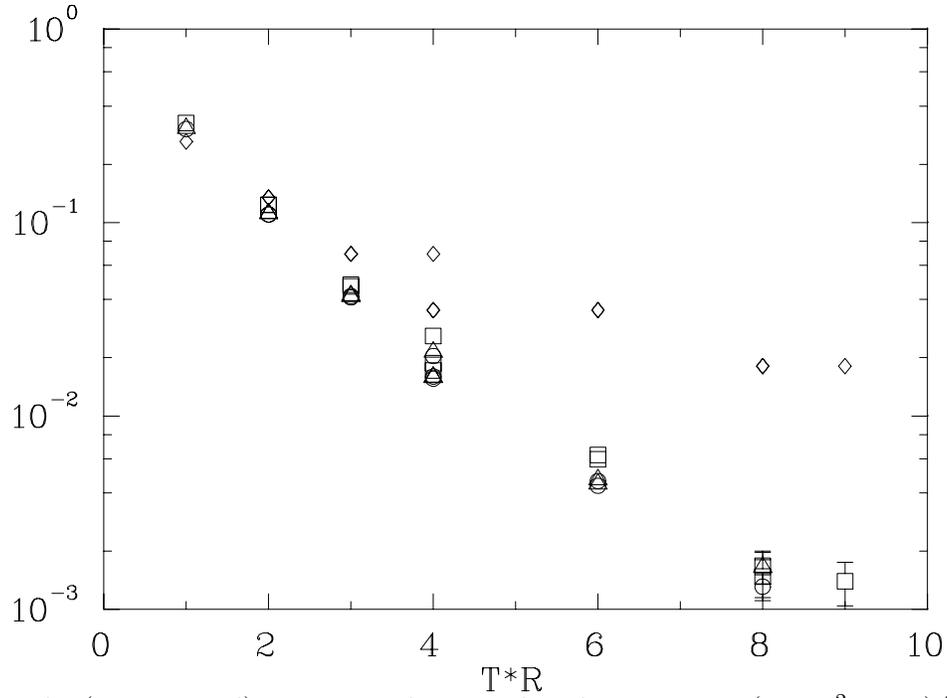,width=.75\textwidth,angle=90}}
\caption[dummy]{ The (unprojected) invariant adjoint Wilson loop $W_{j=1}$ = 
$(4<w_0^2>-1)/3$ (o)
versus the neutral adjoint Wilson loop $(2<w_0^2+w_3^2>-1)$
in F12 ($\triangle$) and MA projection ($\Box$). The approximation,
Eq.~(\ref{exp_sq}), is also shown ($\diamond$).}
\label{n1}
\end{center}
\end{figure}

Let us now use the techniques developed in the previous sections to address 
the adjoint Wilson loop. Using Eq.~(\ref{w03}) we find
\bea\label{squares} 
w_0^2\pm w_3^2 
      &=&  (\cos^2\phi)^{2L}\Bigl(\cos^2\theta_L \pm \sin^2\theta_L\Bigr) \nnn\\
      &+&  (\sin^2\phi)^{2L}\Bigl(\cos^2\chi_L \pm \sin^2\chi_L\Bigr)     \nnn\\
      &+& 2(\cos\phi)^{2L}\sum_m (\cos\phi)^{2m}(\sin\phi)^{2(L-m)}
                     \sum_n s_n \cos(\O_n\mp\theta_L)            \nnn\\
      &+& 2(\sin\phi)^{2L}\sum_m (\cos\phi)^{2m}(\sin\phi)^{2(L-m)}
                   \sum_n s_n \cos(\O_n\mp\chi_L)                \nnn\\
      &+& 2(\cos\phi)^{2L} (\sin\phi)^{2L} \Bigl(\cos(\theta_L\mp\chi_L)\Bigr)\nnn\\
      &+&  \sum_{m,m'} (\cos^2\phi)^{(m+m')}(\sin^2\phi)^{(2L-m-m')}
                  \sum_{n,n'} s_n s_{n'} \cos(\O_n\mp\O_n') \ .
\eea
The first issue is to understand why the neutral part of the Wilson loop
does not fluctuate around $1/3$ as the diagonal approximation suggests
but rather fluctuates around zero 
even in MA projection~\cite{DD1}. Consider therefore the operator
$(w_0^2 + w_3^2)$. Performing the $\chi$ variable integration in the path integral
by ignoring the small $\chi$-dependent part of the action 
allows only terms with no $\chi$ fields to survive. Since there is no 
$\theta$ dependence at this level we obtain
\bea\label{exp_sq}
<w_0^2+w_3^2> &=& (\cos^2\phi)^{2L} +  (\sin^2\phi)^{2L}
                  + \sum_{m=1}^{L-1} (\cos^2\phi)^{2m}(\sin^2\phi)^{2(L-m)}
                   {2L \choose 2m}              \nnn\\
              &=& {1\over 2} \left[
	\sum_{l=0}^{2L}(\cos^2\phi)^{l}(\sin^2\phi)^{2L-l}
                       {2L \choose l}
                  + \sum_{l=0}^{2L}(\cos^2\phi)^{l}(-\sin^2\phi)^{2L-l}
                       {2L \choose l}\right]  \nnn\\
              &=& {1\over 2} \Bigl[ ( \cos^2\phi + \sin^2\phi)^{2L}
                                   +( \cos^2\phi - \sin^2\phi)^{2L} \Bigr]   \nnn\\ 
              &=& {1\over 2} \Bigl[ 1 + \cos(2\phi)^{2L} \Bigl] \ .
\eea
Thusly, $W^0_{j=1}$ in MA projection fluctuates around $0$ for any value
of $\cos\phi$ not exactly one and {\it not} around $1/3$ as the diagonal 
approximation suggests. This shows that the adjoint diagonal Wilson 
loop is evidently unsuited for testing adjoint abelian dominance 
as is clearly seen in Fig. \ref{www}.  Notice that 
in order to obtain the correct value around which   $W^0_{j=1}$ fluctuates
it is necessary to include $(\sin\phi)$ terms which, although small, are
associated with large degeneracy factors. This is qualitatively different from
the behavior of the fundamental Wilson loop, Eq.~(\ref{pol1}), and the 
behavior of the charged adjoint Wilson loop (Eq.~(\ref{pol2}) below) where only 
($\cos\phi$) terms survive the free $\chi$ integration. Since the
 $\chi$-dependent part of the action
is associated with ($\sin\phi$) factors, consistency requires that the correct
behavior of $W^0_{j=1}$ cannot be predicted unless $\chi$-fluctuations are properly
taken into account by using the $\chi$-dependent part of the action.  
Indeed, from Fig.~\ref{n1} it appears that $W^0_{j=1}$ falls with an area law;
Eq.~(\ref{exp_sq}) fails to describe $W^0_{j=1}$ except for very small loops. 
Physically, this amounts to the fact that
$m=0$ source components interact either via the exchange of charged gluons
(necessitating $<\sin\phi>\ne 0$) or via photons which see an $m=0$ source
``dressed'' with a virtual cloud of charged gluons. In our picture 
this latter mechanism amounts to bringing down from the exponential
powers of the $\chi$ part of the action, which, coupled to $W^0_{j=1}$ will
generate an assortment of perimeter and area terms involving $\theta_L$.
The confining behavior $W^0_{j=1}$ depends on the interplay between these terms.
We conclude that our picture cannot predict a {\it simple} abelian operator 
that generates the string tension
corresponding to the neutral adjoint Wilson loop $W^0_{j=1}$.

\begin{figure}[htb]
\begin{center}
\mbox{\epsfig{file=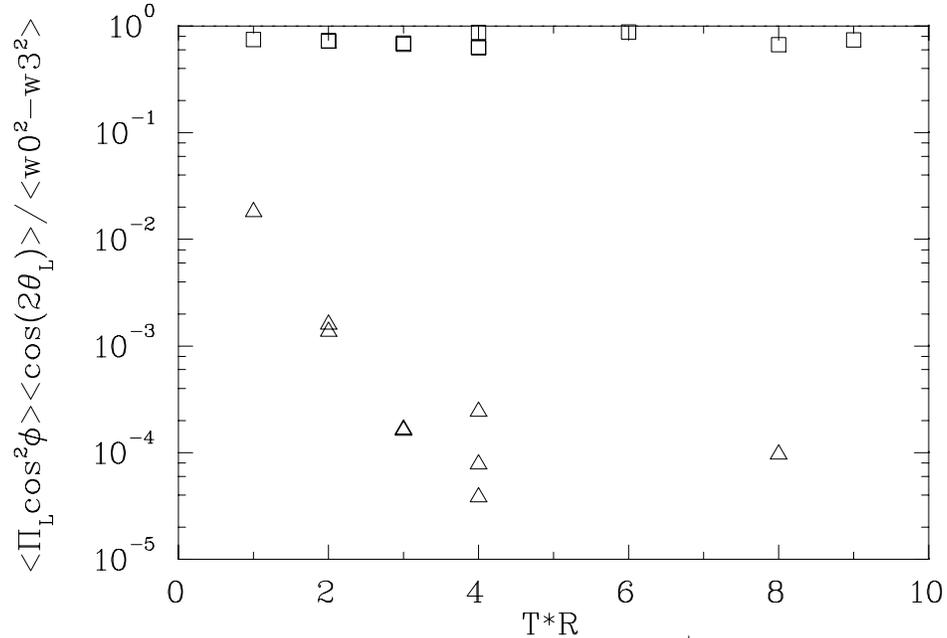,width=.75\textwidth,angle=90}}
\caption[dummy]{ Approximating the charged adjoint Wilson loop $W^{\pm}_{j=1}$
by the factorized product of $<\prod_L\cos^2\phi>$ around the $T\times R$ loop
times the doubly charged abelian Wilson loop $<\cos(2\theta_L)>$.
Results shown in F12 ($\triangle$) and MA ($\Box$) projection.}
\label{ch}
\end{center}
\end{figure}

\begin{figure}[htb]
\begin{center}
\mbox{\epsfig{file=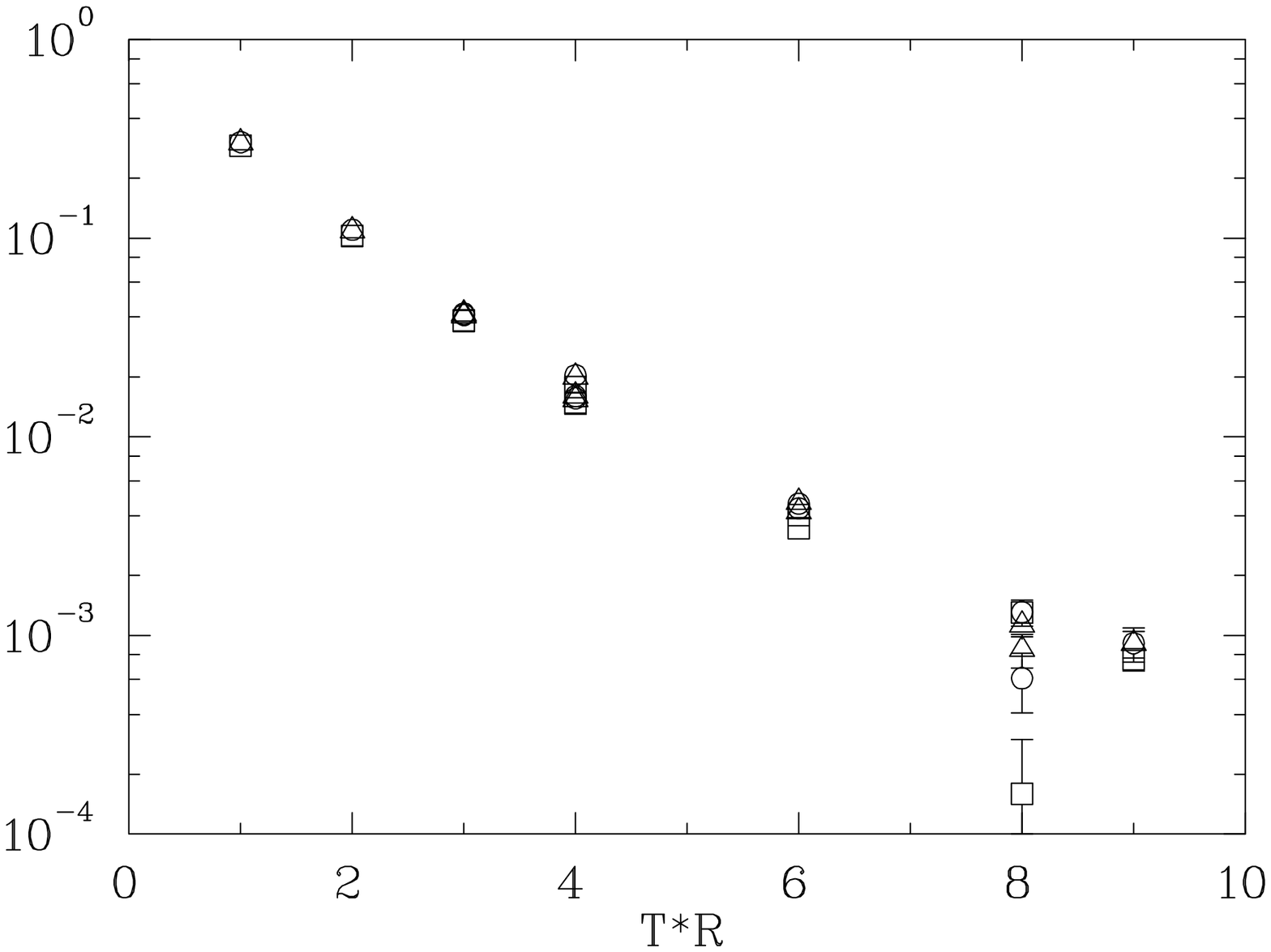,width=.75\textwidth,angle=90}}
\caption[dummy]{ The projection invariant adjoint Wilson loop (o)
compared with the charged adjoint Wilson loop $<w_0^2-w_3^2>$
in F12 ($\triangle$) and MA projection ($\Box$).}
\label{akomh}
\end{center}
\end{figure}

\begin{figure}[htb]
\begin{center}
\mbox{\epsfig{file=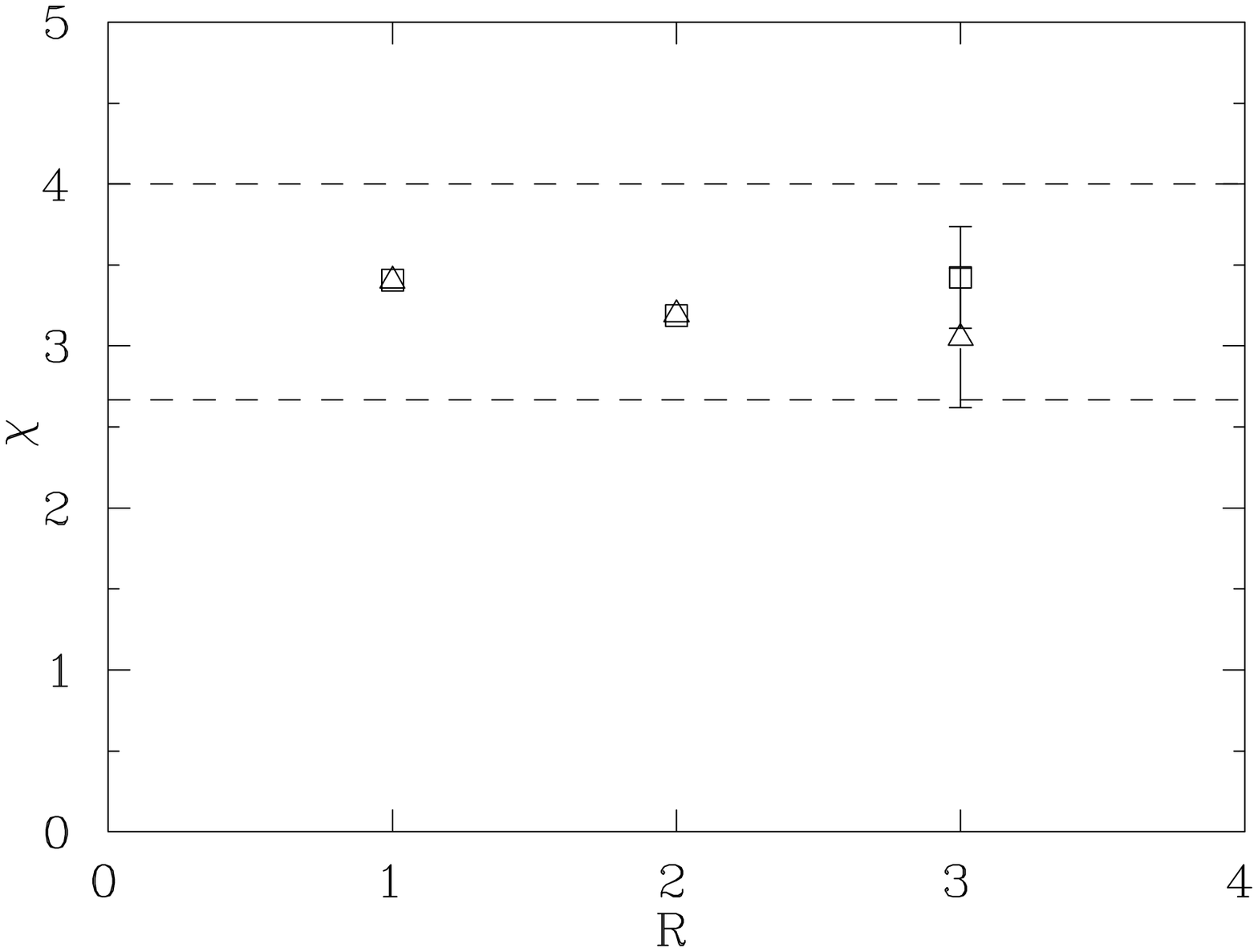,width=.75\textwidth,angle=90}}
\caption[dummy]{	
The ratio $\chi$, Eq.\ (\ref{def_chi}), between the Creutz ratios from
	doubly charged, $W_{n=2}$, and  singly charged, 
        $W_{n=1}$, abelian Wilson loops in  MA projection.
	Results are presented from the $16^4$ run ($\Box$) and 
        the $12^4$ run ($\triangle$). The dashed 
	lines show the Casimir scaling limit of compact QED, $\chi=4$, 
	and the abelian dominance limit,  $\chi=8/3$.}
\label{chir}
\end{center}
\end{figure}

The situation is different for the operator $(w_0^2 - w_3^2)$. Since there 
is no $n,n'$ such that
$\cos(\O_n + \O_n')$ contains no $\chi$, the $\chi$ integration leaves simply
\bea\label{pol2}
<w_0^2-w_3^2>& \approx & (\cos^2\phi)^{2L}\int d[\theta] e^{-[\beta S_\theta + \Delta
S_{FP}] }\cos(2\theta_L) \nnn\\
&=& (\cos^2\phi)^{2L} <\cos(2\theta_L)>_{abel}\ .
\eea
This is the adjoint analog of Eq.~(\ref{pol1}) and is shown in Fig. \ref{ch} to be
a very good approximation in MA projection (only). 
The $(\cos^2\phi)^{2L}$ factor is a perimeter
term as can be seen in  Fig.~\ref{cos2}. The physical interpretation
of this prediction is completely analogous to the fundamental case:
the adjoint source components $m=\pm 1$ are electrically doubly charged; thus,
abelian dominance in this case requires that the string tension
from the doubly charged abelian Wilson loop $<\cos(2\theta_L)>$ 
reproduces the corresponding SU(2) string tension. What is not clear yet 
is the relation between the SU(2)
string tension from $W^{\pm}_{j=1}$ and the string tension from the
gauge-invariant adjoint Wilson loop $W_{j=1}$.
From Figs. \ref{akomh},\ref{n1} can be seen
that $<w_0^2-w_3^2>$ $\approx$ $W_{j=1}$ $\approx$ $<2(w_0^2+w_3^2)-1>$
in both F12 and MA projection. Thus, the full adjoint string
tension and the string tension from the charged adjoint loop $W^{\pm}_{j=1}$
are roughly the same. For example, the (2,2) Creutz ratio from $<w_0^2-w_3^2>$
is $6\%$ higher than the gauge-invariant Creutz ratio, and the one from
$<2(w_0^2+w_3^2)-1>$ is $11\%$ lower. This is in agreement with the results 
in three dimensions~\cite{DD1}.  We can only speculate about the 
origin of these results. A first argument is that it
seems difficult to find string tensions $\sigma_{j=1}$, $\sigma^{\pm}_{j=1}$
and $\sigma^{0}_{j=1}$ such that $\exp(-\sigma_{j=1}\cdot A)$ = 
$a\,\exp(-\sigma^{\pm}_{j=1}\cdot A)$  $+$ $b\,\exp(-\sigma^n_{j=1}\cdot A)$ 
for a range of values for the area $A$ unless $\sigma_{j=1}$ = 
$\sigma^{\pm}_{j=1}$ = $\sigma^{0}_{j=1}$. Another argument is using the
spectral decomposition: although gauge-dependent, $W^{\pm}_{j=1}$ is a
correlator evolving in time with the gauge-invariant Hamiltonian.
The gauge-dependence is manifest only in the ground state to vacuum matrix
element of the operator, which does not affect the area law. If correct, 
this would explain why $\sigma_{j=1}$ = $\sigma^{\pm,AP}_{j=1}$ 
independent of the projection AP. At least what we can do is use 
the equality of the ``full adjoint'' and the ``charged adjoint'' 
string tensions as phenomenological input.
Using $\sigma_{j=1}$ = $\sigma^{\pm}_{j=1}$ in combination with Fig. \ref{ch} 
and Eq.~(\ref{pol2}) suggests that the full adjoint SU(2) string 
tension is given by the doubly charged abelian Wilson loop, 
$W_{n=2}=\cos(2\theta_L)$. In other words,
the observed adjoint string tension arises from the interaction of 
diagonal gluons (photons) with the part of the adjoint source that
is charged (doubly) with respect to the Cartan subgroup. If this is 
indeed verified by simulation it should be regarded as adjoint abelian 
dominance. Notice the difference from the fundamental representation. There, the
Wilson loops from singly charged U(1) links and the one from
diagonal rescaled SU(2) links are identical (c.f. Eqs.~(\ref{e10},\ref{e11})).
In the adjoint case, however, $<\cos(2\theta_L)>$ and the diagonal adjoint
Wilson loop, Eq.~(\ref{e16}),  are proportional up to a constant $1/3$.
Thus, they can not simultaneously lead to an
area law. This difference reflects the fact that fundamental abelian dominance
is oblivious to the difference between MA projection and the diagonal approximation,
whereas adjoint abelian dominance is not. 
If verified by simulations, adjoint 
abelian dominance is a non-trivial result for the following reason:
it is well known that in compact electrodynamics Wilson loops constructed from
U(1) links in representation $n$ (i.e., $n$-ply charged) lead to string
tension scaling with the U(1) Casimir $n^2$~\cite{Banks,TW}. Thus,
for $\beta$ values such that the effective coupling is less than the critical
CEQD coupling, $\beta(\cos\phi)^4 < \beta_c =  1$, and if we ignore the Fadeev-Popov
determinant in the action, we expect
\begin{equation}\label{e22}
	\sigma^{MA}_{n=2} = 4 \;\sigma^{MA}_{n=1} \ ,
\end{equation}
while from Eq.~(\ref{aad3}) adjoint abelian dominance requires 
the relative strength to be $2.7$ rather than $4$. Thus, the non-Wilson like terms
in the effective abelian action that come from the loop expansion of the 
Fadeev-Popov determinant, such as $\prod_P\cos^2\phi\cos(2\theta_P)$ 
terms~\cite{Misha1}, need play a crucial role in bringing this number down from
4 to 2.7. Direct evidence for the presence of such terms in the effective
action has been reported in Ref.~\cite{KenU} by using an inverse microcanonical
demon method to generate the effective abelian action given an ensemble of 
MA-projected QCD configurations. In Fig.\ \ref{chir}
we show results for the ratio
      \begin{equation}\label{def_chi}
	\chi \equiv {\sigma^{AP}_{n=2}/\sigma^{AP}_{n=1}}\label{e24a}\\
	\end{equation}
of doubly to singly charged abelian Creutz ratios in maximal 
abelian (AP=MA) projection from 200 configurations on $16^4$ lattices
and 350 configurations of $12^4$ lattices. For the $R$ values in Fig.\ \ref{chir} the
ratio of adjoint to fundamental SU(2) Creutz ratios is 2.6027(3),
2.553(7) and 2.67(17) at $R$ = 1,2,3, respectively. At $R=1,2$
the singly charged Creutz ratios in MA projection underestimate the 
full SU(2) and therefore $\chi$  is seen in Fig.\ \ref{chir} to
be lower than 4 but above 2.7. Certainly, more calculations are needed
to verify that  $\chi$ $\rightarrow$ 2.7 in MA projection.
The corresponding ratios for in F12 projection are extremely noisy; only $R=1,2$
could be measured with values 2.111(1) and 1.93(50). Assuming a monotonic
decrease with $R$ the results indicate that abelian QCD in F12 projection
is rather different than compact QED. Wensley~\cite{Roy}
has measured the string tension from doubly and singly charged 
abelian loops using monopoles alone. At $\beta$ = 2.5 he finds 
$\sigma_{j=1/2}$ = 0.033(1) for the unprojected theory, $\sigma^{MA}_{n=1}$ 
= 0.034(1) for the singly charged abelian loops and 0.093(1) for the 
doubly charged $\sigma^{MA}_{n=2}$, thusly yielding a ratio 2.8(3). 
The trend seen in Fig.\ \ref{chir} is in qualitative agreement with these
results.

\begin{figure}[htb]
\begin{center}
\mbox{\epsfig{file=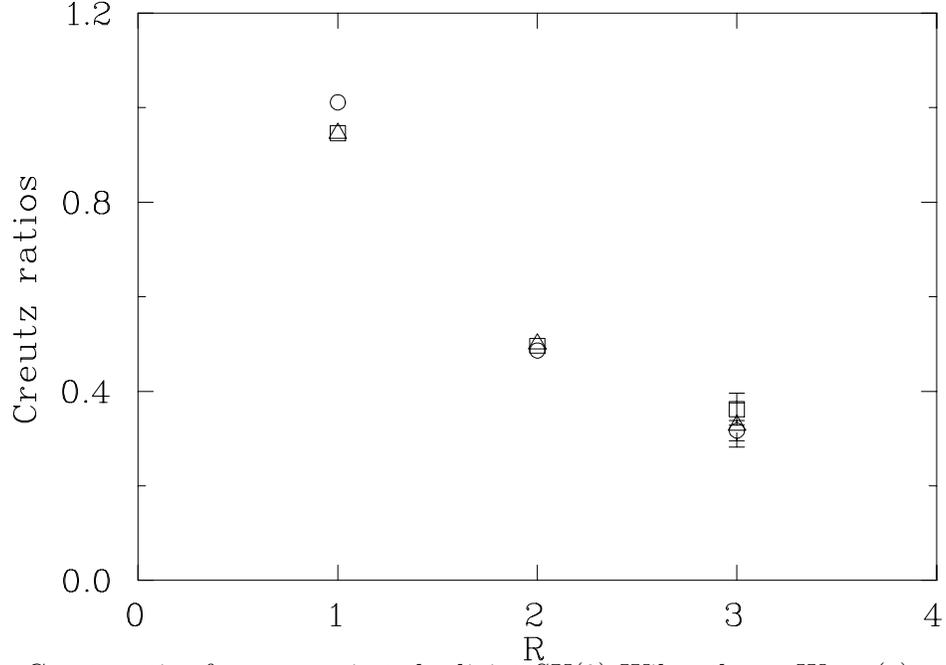,width=.75\textwidth,angle=90}}
\caption[dummy]{Creutz ratios from
        unprojected adjoint SU(2) Wilson loops $W_{j=1}$ (o)
        versus Creutz ratios from doubly charged
        abelian Wilson loops  $W_{n=2}$ in MA projection 
	from the $16^4$ run ($\Box$) and the $12^4$ run ($\triangle$).}
\label{dad}
\end{center}
\end{figure}

{\it Direct} numerical evidence for 
adjoint abelian dominance is shown in Fig. \ref{dad} where Creutz ratios
from full SU(2) Wilson loops are compared to  Creutz ratios from 
doubly charged abelian loops. As we have remarked earlier on, the lack
of variance reduction and smearing techniques makes the calculation
of doubly charged abelian Wilson loops very difficult. Thus, only results up
to $R=3$ are shown. These results seem to support adjoint abelian dominance but 
certainly more calculations (demanding resources unavailable to us) are 
needed before definitive conclusions may be drawn. The F12 projection
doubly charged Creutz ratios  (not shown in the figure) 
behave similarly to the singly charged ones in Fig.\ \ref{fad}: they are 
more difficult to measure but consistently higher than the unprojected 
SU(2) ones.
 
The last topic that should be discussed is the physics of charge screening
in abelian projected QCD with adjoint sources. In the diagonal 
approximation picture screening comes about automatically since $W^d_{j=1}$
does not support a string tension, as we have already discussed.
However, with respect to the conjecture
that abelian dominance can be understood in the context of doubly charged 
abelian Wilson loops, there appears to be an important caveat: once the doubly
charged abelain Wilson loops develop a string tension there seems to be no way 
that this abelian flux-tube can break, since the photons are neutral themselves
and cannot couple with the charged sources to form electrically neutral states
(the analog of ``gluelumps''). Of course one may argue that the off-diagonal 
gluons $A_{\pm}$ which carry double charge can now form electrically neutral 
states with the charge-two sources thus screening the abelian potential 
obtained from doubly charged Wilson loops. It is more subtle though to 
address this question in the context of an effective abelian theory 
 that results from the integration of the charged gluons ($\chi$). This theory
(c.f. Eq.~(\ref{pol2})) contains just photons (and monopoles). 
What happens is probably the
following~\cite{KKK}: the loop expansion of the Fadeev-Popov determinant
leads to terms in the effective $\theta$  action that contain doubly charged 
plaquettes of size $L\approx R_c$ where $R_c$ corresponds to the critical 
distance where gluelumps screen the adjoint sources in full SU(2). Due to the 
numerical difficulties the presence of these terms in the action 
has not been established in Ref.\ \cite{KenU} 
but they are presumably there. When the Wilson loop is large enough to 
``accommodate'' such terms one can use them (only a
small number is necessary!) to ``patch'' the area of the Wilson loop~\cite{RRR}.
As a result the Wilson loop does not fall like the area any more (as would be 
the case with only $1\times 1$ plaquettes in the action) and a crossover to
perimeter law falloff occurs.

\section{summary}

In abelian projected QCD monopoles, contained
in the diagonal gluon (photon) fields $\theta$ are supposed to be responsible
for confinement, by squeezing the abelian flux between electrically charged
sources inside Abrikosov flux-tubes. Thus, a demonstration that abelian Wilson
loops reproduce the full QCD string tension is crucial evidence in favor 
of this mechanism. The maximal abelian projection is the only projection in which
this property of ``abelian dominance'' has been observed, and for quarks 
transforming in the fundamental representation. In this work a
framework for describing the dynamics of abelian projected QCD in the maximal
abelian projection is discussed. Specifically, it is shown that in maximal abelian QCD
\begin{enumerate}
\item[(I)] Gluons are mostly diagonal: at $\beta=2.4$ $<\cos\phi>\simeq 0.93$. 
          The magnitude of off-diagonal gluon 
          fields $<\sin\phi>$ is small and does not fluctuate. Thus,
           $(\cos\phi)$ terms factorize in expectation values. 
\item[(II)] In commensurate with (I) the SU(2) action is by $90\%$ 
          (at $\beta=2.4$) accounted for by
         a compact QED action with effective coupling $\beta(\cos\phi)^4$.
         Thus, the phases $\chi$ of off-diagonal gluons are basically random. 
\item[(III)] Using (I) and (II) to integrate over the (free) $\chi$ variables
             implies that
	\begin{itemize}
\item[(a)]  fundamental SU(2) Wilson loop and singly charged abelian Wilson loop
	expectation values differ by a perimeter term, $(\cos\phi)^{2L}$. 
        Thus, they generate the
        same string tension, which is the familiar abelian dominance for
        fundamental representation quarks.

\item[(b)]            in analogy with [a], 
            doubly charged abelian Wilson loops must generate the
           same string tension as charged (i.e., $m=\pm 1$)
           adjoint SU(2) Wilson loops. The perimeter term in this case
           is $(\cos^2\phi)^{2L}$.

\item[(c)] Adjoint Wilson loops constructed from 
           rescaled SU(2) links should not be used for testing adjoint
           abelian dominance because they fluctuate around $1/3$, in {\it any}
	   abelian projection, and therefore lead to a vanishing abelian
           potential.
             
\item[(d)] the neutral (i.e., $m=0$) adjoint Wilson loop fluctuates
          around 0, not around $1/3$ as ignoring the off-diagonal gluons would
          predict, albeit with a perimeter law falloff. 
            However, $\chi$-integration  cannot be carried out consistently
           unless the $\chi$-dependent part of the action is also included.
           Thus, the abelian operator that generates the string tension of 
           the  neutral adjoint Wilson loop cannot be predicted in a simple way.
\end{itemize}
\item[(IV)]            
           Numerically it is then found that charged Wilson loops 
           have approximately the same, Casimir scaling, string 
           tension as the gauge invariant full adjoint Wilson loop.
           Using this in conjunction with [b] suggests that doubly charged 
           abelian Wilson loops generate the full SU(2) string tension.
           This is verified numerically. 
\item[(V)]
      The ratio of double to singly charged Wilson
        loops in MA projection is below 4:1
       which would be the Casimir scaling limit if the effective U(1) theory after
	the abelian projection is simply compact QED. In MA projection the ratio tends 
	towards the $2.7$ limit that fundamental and abelian dominance, together
	with Casimir scaling  for the unprojected SU(2) theory, would require. 
	In F12 projection the ratio lies below 2 which is an indirect indication that 
	abelian QCD in F12 projection is less close to
	CQED than abelian QCD in MA projection.
\end{enumerate}
Before definitive conclusions can be drawn, two points need to be clarified.
Firstly, the relation between $W^{\pm}_{j=1}$ and $W_{j=1}$ must be
understood. Although we have presented an argument suggesting they should
generate the same Creutz ratios, we feel that before this issue is completely resolved
it is not clear whether our results confirm Eq.~(\ref{aad1}) or only Eq.~(\ref{pol2}).
Secondly, our results should certainly be confirmed at larger interquark 
separations, $5a\le R\le 7a$ at this $\beta=2.4$ value,
where an non-vanishing, roughly Casimir scaling, adjoint string tension is observed.
 Assuming that they survive this test
what is their implication for the abelian projection picture? Clearly, if
one sticks to the notion that the off-diagonal gluons must play no role
in the abelian projection dynamics then the abelian projection breaks down,
as has been advocated in Refs.\ \cite{DD2,DGH,DD1}. The interpretation emerging
from this study is somehow different: off-diagonal gluons cannot be neglected, but
still, abelian degrees of freedom, i.e., electrically charged source components and
diagonal gluons (photons), ``dominate'', in the sense that they generate 
the full SU(2) string tension, in both fundamental and adjoint 
representations. In that respect the MA projection appears to be uniquely
successful: it effects what all these studies have set out to accomplish,
namely, to identify a subset of degrees of freedom  in terms of which
the nonperturbative aspects of QCD can be understood. 
The success of the MA projection in reproducing the Casimir scaling
behavior of the adjoint string tension provides further support to the idea that 
confinement arises due to  monopoles contained in the abelian photon 
fields. 
 
\acknowledgments
The author wishes to thank Ken Yee, Maxim Chernodub, Jan Smit, Richard Woloshyn,
and, especially, Mikhail Polikarpov for fruitful discussions and suggestions.
This work is supported
by Human Capital and Mobility Fellowship ERBCHBICT941430.

\vfill\eject

 

\begin{table}[hbt]
\setlength{\tabcolsep}{.15cm}
\renewcommand{\arraystretch}{1.8}
  \caption{Summary of various Wilson loop operators considered in this work
   and the corresponding  string tensions. Notation is as in the text: 
   $j$ labels the SU(2) and $n$ the U(1) (abelian) representations; 
   $d$ denotes 
   diagonal rescaled SU(2) links and dSU(2) the expectations (not
   measurements) in the limit where the ``diagonal approximation'' is exact,
   corresponding to $w_0^2+w_3^2$ $\equiv$ 1 at the operator level
   (equivalently, $\cos\phi\equiv 1$). These are to be contrasted with 
   the results in MA and F12 projection abelian QCD (last two columns). 
   Abelian dominance (AD) is tested with operators 
   expressed in terms of abelian link variables. Thus, when possible, operators
   are presented in terms of abelian Wilson loops, with $\theta_L\equiv$
   $\cos(\sum_{i\in L} \theta_i)$. 
  }
\begin{tabular}{cccccc}\hline
                 \multicolumn{1}{c}{$\mbox{notation}$}
                 & \multicolumn{1}{c}{$\mbox{interpretation}$}
                 & \multicolumn{1}{c}{$\mbox{definition}$}
                 & \multicolumn{1}{c}{$\mbox{dSU(2)}$}
                 & \multicolumn{1}{c}{$\mbox{MA}$}
                 & \multicolumn{1}{c}{$\mbox{F12}$}\\ \hline
$W_{j=1/2}$           
& $\mbox{SU(2), fund.}$ & $ w_0 $  & $\sigma$ &  $\sigma$ & $\sigma$\\
$W^d_{j=1/2}\equiv W_{n=1}$ & $\mbox{U(1), n=1}$ &  $\cos\theta_L $ & $\sigma$ & 
   $\approx\sigma\;(\mbox{AD})$ & $ > \sigma$\\
$W_{j=1}$ & $\mbox{SU(2), adj.}$ & ${4 \over 3}w_0^2-{1\over 3}$  & $0$ 
	& ${8\over 3}\sigma$ & ${8\over 3}\sigma$\\
$W^{\pm}_{j=1}$ & $\mbox{adj.}, m=\pm 1$ & $ {2\over 3}(w_0^2-w_3^2)$ &
        $ $ & $\approx{8\over 3}\sigma$ & $\approx {8\over 3}\sigma$ \\
$W^{0}_{j=1}$ & $\mbox{adj.}, m=0$ & ${2\over 3}(w_0^2+w_3^2)-{1\over 3}$ &
        $0$ & $\approx {8\over 3}\sigma$ & $\approx {8\over 3}\sigma$ \\
$W^{d}_{j=1}$ & $\mbox{adj., diag. }$  & ${2\over3}\cos(2\theta_L)+{1\over 3}$ &
        $0$ & $0$ & $0$\\
$W_{n=2}$ & $\mbox{$U(1)$, n=2}$  & $\cos(2\theta_L)$ &
        $ $ & $\approx{ 8\over 3}\sigma \; (\mbox{AD})$ & $ >{8\over 3}\sigma$\\
\hline
\end{tabular}
\end{table}

\noindent


\begin{references}
\bibitem{term} We use the term ``projection'' rather than  ``gauge'' 
               since the gauge is only partially fixed.
\bibitem{infi} An SU(2) element parametrized by $w_a$, $a\in\{1,2,3\}$,
has representations ${\cal D}_{j}[w]$ = $\exp(i T_j^a w_a)$, with $T_j^a$
the generators in representation $j$. A quick proof of Eq.\ (\ref{e12}) 
follows by taking $w_a$ infinitesimal and using 
$[T_{1/2}^a]_{bc}=1/2[\sigma_a]_{bc}$
for the fundamental and $[T_{1}^a]_{bc}=-i\epsilon_{abc}$ for the adjoint
SU(2) generators. The proof can be generalized to finite $w_a$.
\bibitem{RRR} The character expansion in strong coupling
	      implies that (depending on their relative coefficient
              in the action) charge 2 terms can dominate~\cite{Zou}
              the expectation value of the doubly charged Wilson loop.
\end{references}
\end{document}